\RequirePackage{amsthm}
\documentclass[sn-mathphys-num]{sn-jnl}

\usepackage{graphicx}%
\usepackage{multirow}%
\usepackage{amsmath,amssymb,amsfonts}%
\usepackage{amsthm}%
\usepackage{mathrsfs}%
\usepackage[title]{appendix}%
\usepackage{xcolor}%
\usepackage{textcomp}%
\usepackage{manyfoot}%
\usepackage{booktabs}%
\usepackage{array}
\usepackage{algorithm}%
\usepackage{algorithmicx}%
\usepackage{algpseudocode}%
\usepackage{listings}%
\usepackage{titlesec}
\titlespacing*{\subsubsection}{0pt}{1pt}{1em} 
\titleformat{\subsubsection}[runin]{\normalfont\normalsize\bfseries}{\thesubsubsection}{1em}{}

\definecolor{codegray}{rgb}{0.95,0.95,0.95}
\definecolor{darkgreen}{rgb}{0,0.6,0}
\definecolor{darkblue}{rgb}{0,0,0.7}

\lstdefinestyle{pythonprompt}{
    language=Python,
    basicstyle=\ttfamily\small,
    keywordstyle=\color{keywordblue}\bfseries,
    stringstyle=\color{darkgreen},
    commentstyle=\itshape\color{darkgreen},
    frame=none,
    showstringspaces=false,
    breaklines=true,
    breakindent=0pt,
    tabsize=4,
    captionpos=b,
    xleftmargin=0pt,
    xrightmargin=0pt,
    aboveskip=1em,
    belowskip=1em
}

\theoremstyle{thmstyleone}%
\theoremstyle{thmstyletwo}%

\theoremstyle{thmstylethree}%
\begin{document}

\title[Article Title]{OptiChat: Bridging Optimization Models and Practitioners with Large Language Models}

\author[1]{\fnm{Hao} \sur{Chen}}\email{chen4433@purdue.edu}

\author[1]{\fnm{Gonzalo Esteban} \sur{Constante-Flores}}\email{geconsta@purdue.edu}

\author[1]{\fnm{Krishna Sri Ipsit} \sur{Mantri}}\email{mantrik@purdue.edu}

\author[1]{\fnm{Sai Madhukiran} \sur{Kompalli}}\email{skompall@purdue.edu}

\author[1]{\fnm{Akshdeep Singh} \sur{Ahluwalia}}\email{ahluwala@purdue.edu}

\author*[1]{\fnm{Can} \sur{Li}}\email{canli@purdue.edu}

\affil*[1]{\orgdiv{Davidson School of Chemical Engineering}, \orgname{Purdue University}, \orgaddress{\city{West Lafayette}, \postcode{47907}, \state{IN}, \country{USA}}}




\abstract{Optimization models have been applied to solve a wide variety of decision-making problems. These models are usually developed by optimization experts but are used by practitioners without optimization expertise in various application domains. As a result, practitioners often struggle to interact with and draw useful conclusions from optimization models independently. To fill this gap, we introduce OptiChat, a natural language dialogue system designed to help practitioners interpret model formulation, diagnose infeasibility, analyze sensitivity, retrieve information, evaluate modifications, and provide counterfactual explanations. By augmenting large language models (LLMs) with functional calls and code generation tailored for optimization models, we enable seamless interaction and minimize the risk of hallucinations in OptiChat. We develop a new dataset to evaluate OptiChat's performance in explaining optimization models. Experiments demonstrate that OptiChat effectively delivers autonomous, accurate, and instant responses. These findings highlight the potential of LLMs to bridge the gap between optimization models and practitioners in the real-world decision-making process.}




\maketitle

\section{Introduction}\label{sec:introduction}
Significant progress in large language models (LLMs) has been witnessed in recent years \citep{OpenAI2023, Touvron2023}, with applications emerging in chemistry \citep{Andres2024, Jablonka2024, Boiko2023, DrugLi2023}, biology \citep{Lin2023, Luo2022}, healthcare \citep{thawkar2023, Sallam2023}, finance \citep{Dowling2023, wu2023, lopezlira2024}, manufacturing \citep{Wang2023, Badini2023}, supply chain management \citep{LiMicrosoft2023}, and optimization modeling \citep{ahmaditeshnizi22024, ahmaditeshnizi2024, ahmaditeshnizi2023, xiao2024, tang2024, ramamonjison2023}. This widespread applicability highlights the capability of LLMs to comprehend and process natural language across diverse domains. Leveraging the versatility of LLMs, a recent study introduced an interactive dialogue system, \textit{TalktoModel} \citep{Slack2023}, designed to assist practitioners in understanding complex machine learning models from various application domains by integrating LLMs with explainable AI (XAI) techniques \citep{Ribeiro2016, Lundberg2017, Selvaraju2017, karimi2020}. Case studies demonstrated that the interactive dialogue system of \textit{TalktoModel} significantly enhanced healthcare workers' understanding of an ML-based disease prediction model.

In addition to machine learning models, optimization models are another type of complex model widely used across various fields, such as engineering, economics, healthcare, and manufacturing \citep{Rardin2016}. These models formulate real-world decision-making problems into optimizing an objective under a set of constraints (e.g., budgets and operational limits), where each decision variable has a well-defined physical meaning. For example, a supply chain optimization model aims at determining the most economically efficient production levels and product distribution while meeting customer demands. While optimization models are built on structured and interpretable mathematical formulations that are accessible to experts, practitioners face significant challenges in understanding the abstract formulations and often perceive them as black boxes. This challenge becomes even more pronounced when the solutions suggested by these models deviate from familiar heuristics, leaving practitioners uncertain whether to trust them.  To address these challenges, the optimization community has also developed explanation techniques analogous to those in XAI. Since the 1980s, the community has developed expert systems \citep{Greenberg1983,Greenberg1987}, interpretable decision rules \citep{Bertsimas2021TheOptimization, Goerigk2023AModels, Lumbreras2024ExplainingLearning}, argumentation-based methods \citep{Cyras2021ScheduleScheduling,Cyras2019ArgumentationScheduling,Collins2019TowardsPlanning}, and counterfactual explanations \citep{Forel2023ExplainableAgain}. 

However, the effective use of these explanation techniques still requires substantial optimization expertise. It can still be challenging for the end users of the optimization models, such as logistics coordinators, to independently comprehend the model's outcomes, reconcile results that conflict with their own experiences, or conduct further analysis. This limitation imposes a significant burden on the optimization experts tasked with communicating the results to practitioners, slows decision-making processes, and prohibits the dissemination of optimization models in areas where optimization experts are inaccessible. 

\begin{figure}[ht]
\centering
\includegraphics[width=0.7\linewidth]{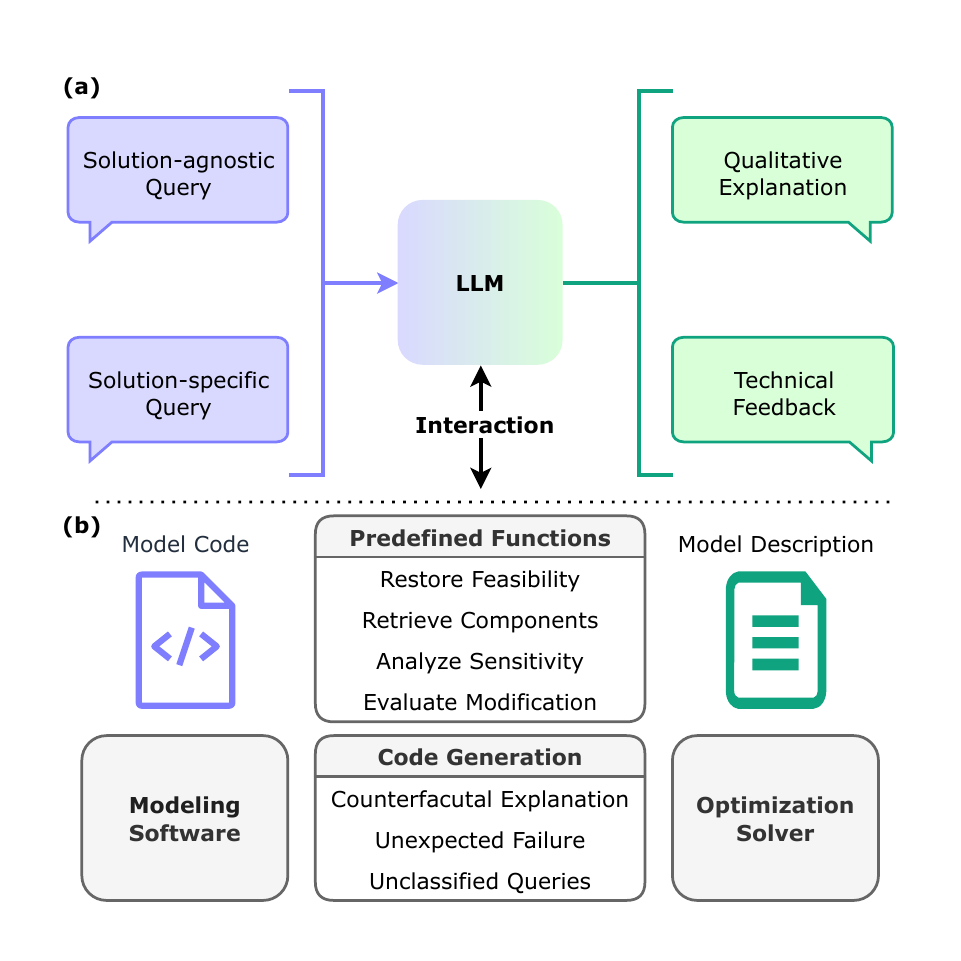}
\caption{Overview of OptiChat. Inputs in blue, outputs in green, and tools in grey. (a) User interface. During a conversation, the solution-agnostic query is answered directly through the LLM's in-context reasoning. The solution-specific query is addressed by interacting with several optimization tools and the model. (b) Backend system. The model code is preprocessed into a natural language description, both of which are accessible to the LLM and the optimization tools. The optimization tools include an optimization solver, an algebraic modeling language, predefined functions, and code generation.} \label{fig:OptiChatOverview}
\end{figure}

To address this limitation, a natural approach is to leverage LLMs for explaining optimization models, given their versatility in natural language-based tasks. However, the hallucination of existing LLMs is concerning because it can produce unfounded or inaccurate explanations \citep{Ji2023}. One solution proposed by \cite{LiMicrosoft2023} is to rely entirely on code generation to handle all queries for a supply chain optimization model. Alternatively, our previous work \citep{OptiChat2023} augments LLMs with predefined functions, but this system is specifically designed for diagnosing infeasible optimization models. Moreover, there is currently no general-purpose platform tailored for explaining optimization models using natural language, nor are there open-source datasets available for systematically evaluating such systems.

\begin{figure}[t]
\centering
\includegraphics{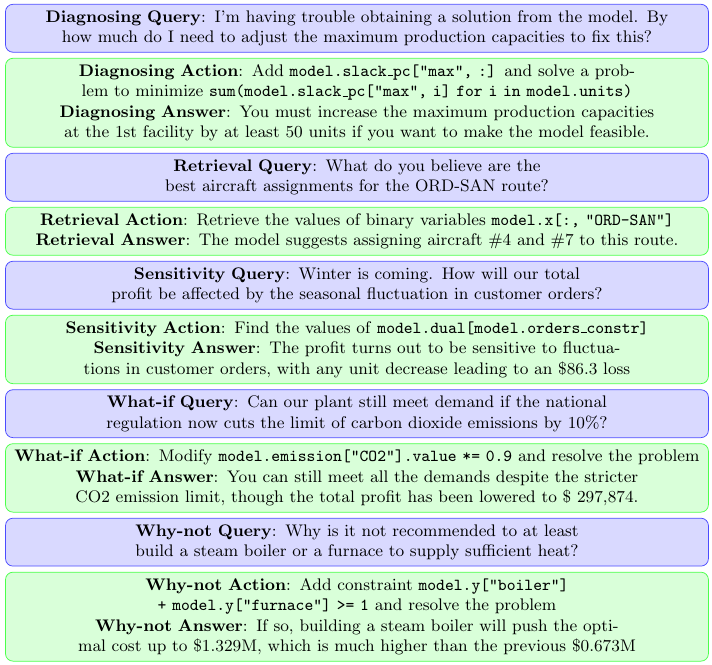}
\caption{Illustrative examples of five different types of queries, the proposed actions, and answers. The proposed actions represent the strategies that generate the relevant explanations, implemented in OptiChat through predefined functions or code generation. The answers are concise summaries of the responses produced by OptiChat.} \label{fig:QuestionClass}
\end{figure}

In view of these challenges, we introduce OptiChat, an LLM-assisted system designed to not only interpret optimization models in natural language but also provide post hoc explanations through interactive dialogues after the optimization model is solved. Common types of user queries can be classified and systematically addressed by OptiChat. To achieve this, we combine various predefined functions and code generation to draw faithful and trustworthy conclusions from optimization models. We also curate a comprehensive dataset to evaluate OptiChat, supporting further research in this area.

As illustrated in Figure \ref{fig:OptiChatOverview}, the input to OptiChat is a well-written code script corresponding to an optimization model in the Pyomo/Python \citep{Hart2011} algebraic modeling language, developed by an optimization expert. OptiChat initially processes the code to generate a coherent and easy-to-understand description to assist practitioners in understanding the problem context. After that, practitioners can ask queries of interest in an interactive dialogue. We survey the queries gathered from practitioners and classify them into five categories: diagnosing, retrieval, sensitivity, what-if, and why-not queries (see illustrative examples in Figure \ref{fig:QuestionClass}). Diagnosing queries focus on fixing infeasibility issues. Retrieval queries seek to extract relevant model data. Sensitivity queries measure the impact of parameter uncertainties on the optimal objective value. What-if queries evaluate significant changes in input parameters specified by users. Lastly, why-not queries investigate the counterfactual scenarios suggested by users. These are termed as \textit{solution-specific queries}, meaning OptiChat must derive a solution from the model to provide an accurate explanation. Underneath, explanation strategies for each query category are either implemented as predefined functions or generated as code by the LLM, which are then executed using an optimization solver, such as Gurobi. These two approaches are tailored to achieve high correctness rates in answering queries. Besides the \textit{solution-specific queries}, the users can also seek qualitative explanations from the LLM. These queries, termed as \textit{solution-agnostic queries}, are contextualized within the LLM's memory using the input optimization model and the chat history to ensure the quality of the answers. 

\section{Related Works}
We give a brief review of recent applications of large language models in operations research. We divide the works into three different categories which correspond to the typical workflow of developing and adopting an optimization model. (1) Use LLM to formulate optimization models based on the problem statement. (2) Use LLM to help develop a new optimization algorithm (3) Use LLM to facilitate interaction with with the user. Our work belongs to the third category. 

\textbf{Formulating optimization models} OptiMUS \citep{ahmaditeshnizi22024,ahmaditeshnizi2023} is a multi-agent LLM-based system designed to formulate and solve (mixed-integer) linear programming problems from natural language descriptions. The system can develop mathematical models, write and debug solver code, evaluate generated solutions, and iteratively improve model and code efficiency and correctness based on these evaluations. \cite{huang2025orlm} proposed ORLM (Operations Research Language Model), a novel approach to OR modeling that relies on fine-tuning a semi-synthetic dataset rather than using prompt engineering or agentic models. \cite{xiao2024} introduced a multi-agent cooperative framework called Chain of Experts (CoE) for automated OR problem modeling. \cite{astorga2024autoformulation} developed a Monte Carlo Tree Search (MCTS)-based approach that decomposes the modeling process into sequential stages (e.g., variables, objectives, constraints) and uses MCTS to explore plausible model structures. \cite{yang2024optibench} presented OptiBench, a benchmark suite for end-to-end optimization problem solving with human-readable inputs and outputs.

\textbf{Develop novel optimization algorithms.} Another line of work focuses on the development of novel OR algorithms. \textit{FunSearch} \citep{funsearch2024}, which explores the function space, is an evolutionary procedure that pairs a pretrained large language model (LLM) with a systematic evaluator. It has been successfully applied to enhance online bin packing heuristics by evolving the heuristics generated by the LLM. A genetic programming algorithm is employed to balance exploitation and exploration within the database of LLM-generated programs. Building on the pioneering efforts of \textit{FunSearch}, several subsequent works have extended similar methodologies to other combinatorial optimization problems \citep{van2024loop}.

\textbf{Facilitating user interactions with optimization models} The final research direction focuses on using LLMs to ease user interactions with optimization models. \cite{Lawless2024IProgramming} demonstrated the use of LLMs to allow users to customize meeting preferences in a constraint programming-based scheduling model. \cite{ju2024globe} designed an LLM-based system for travel planning. \cite{kikuta2024routeexplainer} applied LLMs to explain solutions to vehicle routing problems. Closest to our framework is OptiGuide \citep{LiMicrosoft2023}, which is a multi-agent system for explaining supply chain models by relying entirely on LLM-based code generation to answer user queries. Our main innovation lies in using predefined functions to improve the accuracy and the speed of the agentic framework, as demonstrated through our ablation studies. Compared with the agentic framework of \cite{LiMicrosoft2023}, OptiChat has syntax reminders and Operator agents to coordinate the execution of the predefined functions. We also develop a dataset to evaluate our proposed framework on applications across multiple domains, rather than being restricted to a single use case. This work builds upon our earlier system \citep{OptiChat2023} for diagnosing infeasibility, which was not agentic and did not support queries related to optimality.

\section{Methods}\label{sec:methods}
OptiChat is composed of a user interface, a structured sequence of LLM-based agents, an optimization solver, a modeling software, and several custom-built functionalities. In this work, OptiChat utilizes GPT-4 \citep{OpenAI2023} for the LLM-based agents, Gurobi \citep{Gurobi} as the optimization solver, and Pyomo \citep{Hart2011} as the algebraic modeling language. In this section, we first motivate the functionalities of OptiChat by providing background on common problems faced by practitioners and connecting them with the proposed explanation strategies. We then present the design of our multi-agent framework, outlining the role and sub-task of each agent. Lastly, the exception management for specific queries is discussed. The implementation of OptiChat is avaiable on GitHub: \burl{https://github.com/li-group/OptiChat.git}.

\subsection{Problem Statement And Explanation Strategies}\label{sec:fundamentals}
\subsubsection{Feasible / Infeasible Model Description}
Most practical applications in optimization can be formulated as (mixed-integer) linear programs (MILP/LP):
\begin{equation}\label{eq:MILP}
\begin{aligned}
  \min_{\mathbf{x}} && \mathbf{c}^\top \mathbf{x}\\
  \text{subject to } && \mathbf{A} \mathbf{x} &\le \mathbf{b}\text{,}\\
  && \mathbf{x} &\in \mathbb{Z}^{p} \times \mathbb{R}^{n-p} \text{.}
\end{aligned}
\end{equation}
where decisions to make are denoted as $\mathbf{x}$ that can consist of both integer and continuous variables; $\mathbf{c}$ represents the cost coefficients; the problem is subject to linear constraints $\mathbf{A} \mathbf{x} \le \mathbf{b}$: $\mathbf{A} \in \mathbb{R}^{m \times n}$,  $\mathbf{b} \in \mathbb{R}^m$. However, practitioners often struggle to understand the mathematical formulation. More importantly, optimization models derived from real-world problems can be infeasible at the initial stage. Infeasibility does not refer to code bugs that prohibit models from being executed, but rather to an optimization model in which no solution exists to satisfy all constraints simultaneously. This usually occurs because of overly restrictive constraints and inaccuracies in the underlying data. For example, in an aggressive overselling scenario, an aircraft may lack sufficient seats to accommodate all ticketed passengers. This infeasibility further complicates practitioners' ability to identify flaws in the model. 

An irreducible infeasible subset, IIS, is a minimal set of constraints and/or variable bounds within an optimization model that causes infeasibility, defined by two key properties: (i) the IIS itself is infeasible, and (ii) any proper subset of the IIS is feasible. In other words, we can use the IIS to extract the components in violation and characterize the nature of infeasibility. Different algorithms have been developed to isolate IIS, such as deletion filter \citep{Chinneck1991}, additive method \citep{Tamiz1996}, and hybrid approach \citep{Guieu1999}. Commercial optimization solvers like CPLEX \citep{CPLEX2022} and Gurobi \citep{Gurobi} have implemented variants of these IIS detection algorithms. For a comprehensive review of IIS detection, readers are referred to the monograph \citep{Chinneck2008}. 

\subsubsection{Diagnosing Query}
Restoring feasibility can be approached in various ways, but not all are actionable in the real world. To restore feasibility, practitioners seek strategies that align with their operational priorities. For instance, suppose a model is infeasible to produce a chemical at a 99\% concentration with the current resources available in a chemical plant. Increasing the conversion rate of an existing chemical reaction is generally impractical, while negotiating with business partners to lower product purity requirements is more actionable.

To address this need, one can either (1) recursively remove constraints from the IIS until the model becomes feasible, or (2) add slack variables to the optimization problem and allow for adjustments to input parameters. The second approach is more practical as each constraint reflects an important aspect of the problem and cannot always be removed in practice. In contrast, introducing slack variables into the optimization problem in \eqref{eq:MILP} provides practitioners with a more concrete and actionable plan for feasibility restoration. Mathematically, the following extended problem is solved
\begin{equation}\label{eq:MILPwithSlack}
\begin{aligned}
  \min_{\mathbf{x, \delta A^+, \delta A^-, \delta b^+}} & \quad \sum_{(i,j)\in \mathcal{S}_{A}} \big(\mathbf{\delta A}_{ij}^+ + \mathbf{\delta A}_{ij}^- \big)+ \sum_{i\in\mathcal{S}_{b}}\mathbf{\delta b}_{i}^+ \\
  \text{subject to } & (\mathbf{A} + \mathbf{\delta A^+} - \mathbf{\delta A^-}) \mathbf{x} \le \mathbf{b} + \mathbf{\delta b^+} \text{,}\\
  & \mathbf{x} \in \mathbb{Z}^{p} \times \mathbb{R}^{n-p},\quad  \mathbf{\delta A^+, \delta A^-, \delta b^+}\geq 0 \text{,}
\end{aligned}
\end{equation}
where $\mathbf{\delta A^+}$, $\mathbf{\delta A^-}$, $\mathbf{\delta b^+}$ are nonnegative slack variables. Since all the constraints are inequalities, adding nonnegative slacks $\mathbf{\delta b^+}$ on the right-hand-side relaxes the problem. In contrast, adding the left-hand-side slacks, $\mathbf{\delta A^+}$ and $\mathbf{\delta A^-}$, are less desirable. It is worth noting that because only a subset of parameters can be adjusted in practice, the slack variables differ in dimensionality from the parameters $\mathbf{A}$ and $\mathbf{b}$. 
The objective is to minimize the total perturbation to the original problem by summing up all the slack variables. In principle, different weights could be assigned to different slack variables, representing the cost of perturbing the corresponding parameters. The optimal solution found in \eqref{eq:MILPwithSlack} can be explained as the minimal adjustments needed to restore feasibility. 

It should be noted that adding slack variables to the left-hand-side parameters $\mathbf{A}$ will lead to a product of variables between $\mathbf{\delta A^+}$, $\mathbf{\delta A^-}$ and $\mathbf{x}$ in \eqref{eq:MILPwithSlack}. This results in a non-convex mixed-integer quadratically constrained program (MIQCP), which is often prohibitive to solve. In many situations, left-hand-side parameters represent immutable properties. For example, in the constraint $\mathbf{A} \mathbf{x} \le \mathbf{b}$ where $\mathbf{b}$ is a deadline, $\mathbf{x}$ indicates task status, and $\mathbf{A}$ represents processing times. The processing times are an inherent property of the machines and cannot be changed. When OptiChat detects a request to add slacks to $\mathbf{A}$, it will alert users of this immutability before initiating the MIQCP. 

\subsubsection{Retrieval Query}
An optimization model consists of decision variables, parameters data, constraints, and an objective function. Optimization models developed for practical use are often large in scale, with some components indexed over large sets. For example, a variable that represents the assignment decision of an aircraft can be indexed over hundreds of routes. When practitioners need to review specific data or optimal decisions from the model, it is inefficient to retrieve this information. OptiChat facilitates real-time access to specific model information through natural language.

\subsubsection{Sensitivity Query}
Many optimization models are constructed with incomplete knowledge of problem parameters. For example, electricity prices and customer demands often fluctuate in the market over time and cannot be fixed within the model. The values of these parameters can be predicted using historical data and updated on a rolling basis. It is important for practitioners to evaluate how changes in problem parameters impact the optimal objective value, undertake risk assessment, and devise appropriate management strategies. 

We propose to perform sensitivity analysis based on well-established duality theory in linear programming (LP) \citep{LPbook} where $\mathbf{x}$ does not contain integer variables. In short, the change in the optimal objective value in response to changes in input parameters can be expressed as a value function
\begin{equation}\label{eq:LPsensitivty}
  v(\mathbf{b}) = \min_{\mathbf{x} \in \mathcal{X}} \, \mathbf{c}^\top \mathbf{x} = \mathbf{y^*}^{\top} \mathbf{b} \text{,}
\end{equation}
where $\mathcal{X} = \{\mathbf{x}: \mathbf{A} \mathbf{x} \le \mathbf{b}\}$ and $\mathbf{y^*}^{\top}$ is the optimal solution to the dual problem of \eqref{eq:MILP}. Problem \eqref{eq:MILP} called a primal problem that finds the minimum cost while satisfying all the constraints, whereas the dual problem of \eqref{eq:MILP} aims to seek the tightest lower bound on the optimal primal cost. Denote the original data given in the primal problem as $\mathbf{b^*}$, the expression \eqref{eq:LPsensitivty} indicates that when the perturbed parameters $\mathbf{b}$ are close to $\mathbf{b^*}$, the optimal primal objective value is a linear function of the input parameters with gradient $\mathbf{y^*}$. In other words, it precisely reflects the sensitivity of the optimal objective value in the vicinity of $\mathbf{b^*}$. The access to dual information in LP is supported by most optimization solvers \citep{Gurobi, CPLEX2022, Mosek}. 

\subsubsection{What-if Query}
Although sensitivity analysis can evaluate local changes in parameters, it is not a valid methodology to evaluate larger changes. This limitation arises because the optimal dual solution in \eqref{eq:LPsensitivty} is not constant but depends on parameters. After large parameter perturbations, the dual solution previously found in the original problem no longer accurately reflects the change in optimal objective value. These significant perturbations often occur when a new policy is being established, or the industry is evaluating a new business strategy. For example, practitioners may pose questions such as ``What if we increase the labor force to 35 people?'' and ``What if customer orders are cut by a third?''

In these cases, OptiChat systematically identifies the extent and type of modification from the user's query. 
The model \eqref{eq:MILP} is then revised by updating parameters and constraints accordingly:
\begin{equation}\label{eq:EvaluateModification}
\begin{aligned}
  \min_{\mathbf{x}} && \mathbf{\hat{c}}^\top \mathbf{x}\\
  \text{subject to } && \mathbf{\hat{A}} \mathbf{x} &\le \mathbf{\hat{b}}\text{,}\\
  && \mathbf{x} &\in \mathbb{Z}^{p} \times \mathbb{R}^{n-p} \text{.}
\end{aligned}
\end{equation}
In this revision, $\mathbf{\hat{c}}$, $\mathbf{\hat{A}}$ and $\mathbf{\hat{b}}$ are completed by the interaction between LLMs and our predefined function. Modifications will change the feasible region of the problem, generally leading to a different optimal solution.

\subsubsection{Why-not Query}
Mathematical optimization is a rigorous methodology that searches for the optimal solution. In contrast, practitioners often rely on their experience to make decisions. As a result, it can be challenging to convince business managers or stakeholders to accept the optimal solutions given by an optimization model, especially when these solutions are against their intuitions. For example, one might ask: ``Why not choose supplier 1?'' in a supply chain model, even though the current optimal solution does not include it.

A counterfactual explanation addresses the ``why-not'' queries by examining alternative decisions or outcomes that are not currently realized in the optimal solution. To investigate such a counterfactual scenario, we modify the original model by incorporating an additional set of constraints that force the desired alternative to occur. In this example, if $x_1$ represents the binary decision to select supplier 1, we add the constraint $x_1 = 1$ to enforce this choice.
Once we re-solve the modified problem with this counterfactual constraint in place, we can observe how the objective value and the feasibility of the solution change. For instance, the forced selection of supplier 1 may raise production costs, reduce overall profitability, or even make the problem infeasible. By comparing the new solution to the original one, we gain insight into the trade-offs and underlying reasons that the original optimal solution did not include supplier 1. This counterfactual reasoning thus provides a transparent and actionable explanation of the model’s decision-making process, illustrating which model parameters and constraints are most influential in ruling out the queried alternative. The explanation for the ``why-not'' query relies on code generated by the LLM since the counterfactual can be any arbitrary fact suggested by the practitioners, for which predefined functions do not suffice. 

\subsection{Multi-agent Framework}\label{sec:framework}
\begin{figure}
\includegraphics[width=0.8\linewidth]{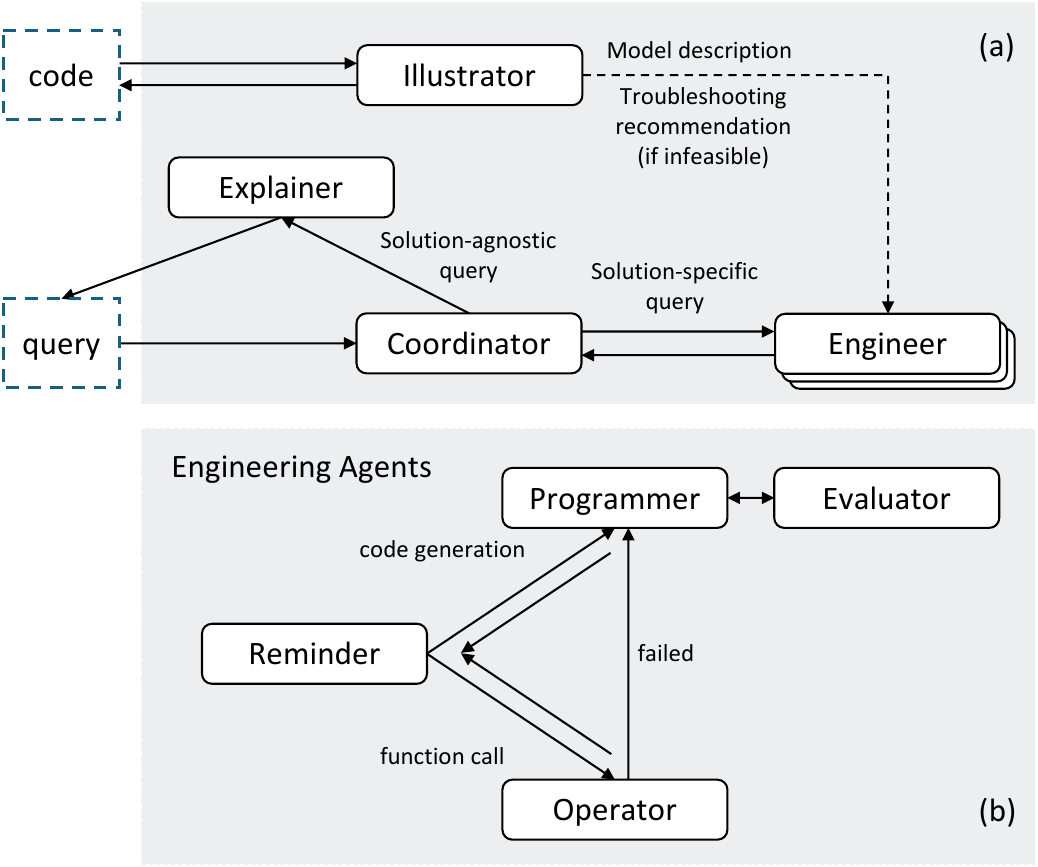}
\caption{Multi-agent framework of OptiChat.
(a) The Pyomo model code is interpreted by the Illustrator agent. Solution-agnostic queries are addressed by the Explainer agent directly. Solution-specific queries are handled by a team of Engineering agents, followed by the Explainer agent. (b) The Engineering agents include the Reminder, Operator, Programmer and Evaluator agents.}\label{fig:MultiAgents}
\end{figure}

Figure \ref{fig:MultiAgents} depicts the workflow of the multi-agent framework. 
Before the interactive dialogue, OptiChat begins with the Illustrator agent, which preprocesses the Pyomo optimization model uploaded by users. During the conversation, the Coordinator routes queries either to the Explainer agent, when they can be directly answered by an LLM, or to the team of Engineering agents when more specialized reasoning is required. The Engineering team operates in a structured order, comprising the following agents: (1) the Operator, who controls all predefined functions; (2) the Programmer and Evaluator agents, responsible for generating code for tasks that cannot be addressed by the predefined functions; and (3) the Reminder agent, who provides supplementary information to enhance the accuracy of the other two. The prompts are shown in Appendix~\ref{sec:prompts}. To illustrate the workflow, a what-if query is provided as an illustrative example in Appendix~\ref{sec:workflow_example}.

\subsubsection{Illustrator}
This agent extracts sets, parameters, variables, constraints, and objectives from the Pyomo optimization model, labeling each component with a natural language description. These descriptions establish a dictionary that maps the physical meanings of model components to their corresponding notations in the source code. Since users may refer to model components using different terms, this preprocessing enables LLMs to accurately identify the referenced component by providing the problem context. Alongside these descriptions, information about components, such as index dimensions and solution status, is also stored to support interaction with optimization tools. Finally, the Illustrator agent describes the model to users in a concise and coherent manner using natural language. If the solution status of a model indicates infeasibility, the agent will also invoke tools to compute the IIS and provide troubleshooting recommendations by interpreting the conflicting constraints.

\subsubsection{Coordinator}
After the Illustrator agent preprocesses a model, users can submit queries to initiate an interactive conversation. The Coordinator agent classifies each query as solution-agnostic or solution-specific. Solution-agnostic queries are solely addressed by the Explainer agent through in-context reasoning, while solution-specific queries require technical feedback from agents in the Engineering team before being forwarded to the Explainer agent. 

\subsubsection{Explainer}
The Explainer agent is responsible for conveying any technical information to practitioners in an understandable manner as the endpoint for all queries. 

\subsubsection{Reminder}
The Reminder agent serves as the entry point for every solution-specific query, guiding LLMs to more accurately determine which function to invoke, which model component to modify, and which specific index within that component to reference. Among these tasks, selecting the correct component index (e.g., a parameter index) is the most error-prone. In contrast, identifying the appropriate function name is generally easier for LLMs, aided by few-shot demonstrations. The model description provided by the Illustrator agent helps the Reminder to pinpoint the relevant component name. The syntax guidance offered by the Reminder agent has been shown in our ablation studies (Section \ref{sec:ablation}) to improve the accuracy in identifying the correct function, component names and indices.

\subsubsection{Operator}
As proposed in Section \ref{sec:fundamentals}, we implement four predefined functions in the Operator agent to address the diagnosing, retrieval, sensitivity, and what-if queries in Figure \ref{fig:QuestionClass}. Informed by the specific syntax guidance, this agent selects the appropriate function and arguments, then invokes the corresponding tool to generate the solution. 

\subsubsection{Programmer And Evaluator}
The development of the Programmer and Evaluator agents is inspired by recent works on applying code generation to optimization \citep{LiMicrosoft2023, ahmaditeshnizi2024}. The code generation capability of LLMs has been widely explored and demonstrated impressive results across various tasks \citep{wu2023autogen}. These agents are designed to address the why-not queries in Figure \ref{fig:QuestionClass}, but also potentially handle unexpected queries beyond the reach of predefined functions. The Programmer and Evaluator are involved in a loop with limited iterations. The Evaluator first executes the code generated by the Programmer, then reviews the terminal outputs and error messages if available. Throughout this process, the code is automatically refined until a bug-free and comprehensive solution is produced. The prompts for why-not queries are designed to guide the LLM in generating additional constraints from the user's counterfactual query, thereby narrowing the LLM's task scope and reducing its code output. 

\subsection{Exceptions Management}\label{sec:exception}
When practitioners interact with OptiChat, the system is robust in handling exceptions caused by the lack of optimization expertise. The sensitivity analysis in OptiChat relies on the strong duality of LP. However, practitioners may request sensitivity analysis concerning left-hand-side parameters $\mathbf{A}$ in LP models or parameters in MILP models. Unlike the sensitivity of $\mathbf{b}$ in \eqref{eq:LPsensitivty}, the dependence of the optimal objective value on left-hand-side parameters $\mathbf{A}$ cannot be determined based on duality theory. When the optimization model is an MILP, the equality in \eqref{eq:LPsensitivty} no longer holds, breaking the connection between the optimal objective value and $\mathbf{b}$. In this case, OptiChat will notify them that sensitivity analysis is not supported and suggest providing specific modifications for evaluation if they still wish to address the same queries. This converts a sensitivity query into a what-if query, which can be addressed in a less restrictive manner. Similarly, when OptiChat detects a request to add slacks to $\mathbf{A}$ in \eqref{eq:MILPwithSlack}, it will issue a warning message to indicate the potential immutability of left-hand-side parameters $\mathbf{A}$ and the increased processing time required to initiate the MIQCP. The model information, such as whether a parameter appears on the left-hand-side, is stored during the preprocessing step. This information will be automatically used to verify the presence of such exceptions in the predefined functions and guide the user accordingly. Furthermore, when unexpected failures occur in predefined functions, the Programmer and the Evaluator agents will be invoked to generate code-based solutions.

\section{Results and Discussion}\label{sec:results}
In this section, the effectiveness of OptiChat is demonstrated in terms of model descriptions and query responses. First, we emphasize the efficiency of OptiChat in generating autonomous model descriptions compared to consulting with optimization experts. In terms of the interactive dialogue, we measure the correctness rates of responses for each query type, providing a quantitative evaluation of OptiChat's accuracy. We also present a qualitative showcase of query responses, illustrating how OptiChat assists practitioners in real-world settings.

We test OptiChat on 24 optimization models written in the Pyomo/Python framework. To demonstrate the versatility of LLMs, the 24 models span various contexts, including supply chain, manufacturing and production, petroleum refinery, industrial scheduling, chemical and process system engineering, transportation, etc. The 24 models are adapted from the GAMS Model Library\footnote{\burl{https://www.gams.com/latest/gamslib_ml/libhtml/index.html}}, sourced from the Pyomo Cookbook by the University of Notre Dame\footnote{\burl{https://jckantor.github.io/ND-Pyomo-Cookbook}} and a public GitHub tutorial\footnote{\burl{https://github.com/hdavid16/RTN-Demo}}, or adapted from an optimization textbook \citep{Rardin2016}. Infeasible variants of these models are created by adjusting the model parameters or introducing additional constraints. A summary of the statistics including the number of variables, constraints, and parameters is shown in Appendix~\ref{sec:datasetsumary}. The size of the models are orders of magnitude larger than those in existing natural language to OR models benchmarks.

OptiChat aims to make optimization models more accessible to a wider audience by enabling seamless interaction between users and the underlying models, thereby reducing the time experts must spend communicating with users. For our experiments, we recruited 29 experts, including graduate students and postdoctoral researchers, each with at least one year of experience in optimization theory or modeling. These experts were tasked with drafting model descriptions and answering follow-up queries.

\subsection{Model Description}
In this study, the expert participants were required to write detailed model descriptions based on the Pyomo scripts of the optimization models. If a model was found to be infeasible, the expert participants were also responsible for diagnosing the source of the infeasibility. Following these tasks, the experts were surveyed to provide time estimates for completing them. Given the differences in expertise among the participants and the varying complexity of the assigned models, we report the most frequently selected time range in the survey. 

On the other hand, OptiChat automatically interprets optimization models in natural language without involving optimization experts, which significantly shortens the time. More importantly, both the experts and OptiChat are augmented with tools to isolate the irreducible infeasible subset (IIS, see details in Section \ref{sec:fundamentals}) for characterizing infeasible optimization models. By isolating the conflicting constraints, optimization experts can gain valuable insight from the IIS analysis. However, it is still time-consuming for them to identify root causes and take corrective actions. In contrast, OptiChat automates the troubleshooting process along with a model description within one minute, as shown in Table \ref{tab:interpretation}. Empirically, the quality of OptiChat's description is observed to be comparable to those provided by experts, which can be attributed to the versatility of LLMs in different problem contexts. Details of the expert survey used to evaluate the model descriptions generated by the LLM are provided in Appendix~\ref{sec:eval_md}.

\begin{table}[ht]
\caption{Model Description Results.} \label{tab:interpretation}
\begin{tabular}{@{\extracolsep\fill}lcc}
\toprule%
 & OptiChat & Expert \\
\cmidrule{2-2}\cmidrule{3-3}%
Model & Time (min) & Time (min)  \\
\midrule
Feasible & $<1$ & $18-40$ \\
Infeasible & $<1$ & $23-55$ \\
\bottomrule
\end{tabular}
\end{table}

In practical applications, the model description in natural language is often already available before the model code is developed, so having the LLM generate such descriptions is not always necessary. Nevertheless, this study serves two important purposes. First, it demonstrates the flexibility of the LLM in explaining why a model is infeasible when an infeasible instance is encountered after model development. Second, by comparing the LLM-generated descriptions with expert answers, we can validate that the outputs produced by the Illustrator agent are sufficiently accurate and reliable to support subsequent query responses.

\subsection{Query Response Assessment}
After reading the autonomous model description, practitioners gain an understanding of the problem context and are prepared to interact with OptiChat by asking queries. Solution-specific queries are addressed in two steps. First, OptiChat either invokes predefined functions or generates code to draw accurate conclusions from the optimization models. Second, OptiChat explains these conclusions in natural language to guarantee their interpretability to practitioners. 

\subsubsection{Query Dataset Development}
We curate a comprehensive test dataset of 172 question-answer pairs to quantitatively evaluate the accuracy of OptiChat in response to the user's query, with each pair developed in the context of a particular optimization model. The queries and answers were drafted by the 29 experts recruited and each verified by multiple authors of the paper. We classify these questions into 5 categories: (1) diagnosing; (2) retrieval; (3) sensitivity; (4) what-if; (5) why-not queries as exemplified in Figure \ref{fig:QuestionClass}. 

The correctness rate of query responses is evaluated using two distinct approaches. For why-not queries that rely on code generation, the dataset provides natural language explanations explicitly containing the optimal objective values of counterfactual models. An LLM is used to compare these ground truth objective values with those produced by OptiChat. If the new optimal objective value computed by OptiChat matches the value specified in the ground truth answer, the LLM marks the response as correct. In rare cases, however, the LLM may generate incorrect counterfactual constraints while still producing a matching objective value. To ensure reliability, we manually review instances marked correct by the LLM to validate both the generated constraints and the resulting answers. In contrast, diagnosing, retrieval, sensitivity, and what-if queries are addressed through predefined functions. These function calls return results that are directly informative to the query. Across all dataset instances, we observe that the Explainer agent does not hallucinate when interpreting function outputs, provided that the correct function and arguments (e.g., component names and indices) are selected. To enable automatic evaluation, ground truth outputs are structured in JSON format. If both the function name and arguments chosen by OptiChat match the ground truth, the corresponding answers are deemed correct.

\subsubsection{Quantitative Assessment of the Query Responses}

Table~\ref{tab:comparison-gpt4o-mini} reports results for four different LLMs: GPT-4o-mini, GPT-4o, GPT-4.1, and o3. The first three are standard pretrained LLMs. o3 is OpenAI's latest reasoning model which features chain-of-thought reasoning at the expense of longer ``thinking'' time.  Among them, o3 achieves the highest average accuracy across all different types of queries. Notably, it achieves significant improved performance on the why-not queries—78.4\% compared to 62.2\% for GPT-4.1 and 54.1\% for GPT-4o—highlighting the value of explicit reasoning capabilities when code generation is required to generate the counterfactual constraints. However, o3 also exhibits higher latency, taking up to 0.9 minutes per response for the ``why-not'' queries, this trade-off appears. In contrast, the model with smallest size, GPT-4o-mini, underperforms across all categories, suggesting limitations in both scale and reasoning depth. Overall, OptiChat delivers strong accuracy (above 80\% in most query types) and fast responses (under one minute), confirming its practicality for interactive use. 

\begin{table}[ht]
\centering
\caption{The accuracy and average response time of the five types of queries using GPT-4o-mini, GPT-4o, GPT-4.1, and o3. The best-performing accuracy for each type of model is highlighted in bold.}\label{tab:comparison-gpt4o-mini}
\begin{tabular}{@{\extracolsep\fill}lcccc|cccc}
\toprule
& \multicolumn{4}{c}{Accuracy (\%)} & \multicolumn{4}{c}{Time (min)} \\
\cmidrule{2-5} \cmidrule{6-9}
Query & GPT-4o-mini & GPT-4o & GPT-4.1 & o3 
      & GPT-4o-mini & GPT-4o & GPT-4.1 & o3 \\
\midrule
Diagnosing   & $53.8$ & $84.6$ & \textbf{89.7} & \textbf{89.7} & $0.1$ & $0.2$ & $0.1$ & $0.6$ \\
Retrieval    & $66.7$ & $92.3$ & $94.9$ & \textbf{97.4} & $0.1$ & $0.1$ & $0.1$ & $0.3$ \\
Sensitivity  & $72.2$ & \textbf{94.4} & \textbf{94.4} & \textbf{94.4} & $0.1$ & $0.2$ & $0.1$ & $0.5$ \\
What-if      & $64.1$ & \textbf{94.9} & $84.6$ & $87.2$ & $0.1$ & $0.2$ & $0.1$ & $0.4$ \\
Why-not      & $62.2$ & $54.1$ & $62.2$ & \textbf{78.4} & $0.3$ & $0.4$ & $0.3$ & $0.9$ \\
Total      & $62.8$ & $83.1$ & $84.3$ & \textbf{88.9} & $0.2$ & $0.2$ & $0.2$ & $0.5$ \\
\bottomrule
\end{tabular}
\end{table}

The high accuracy of OptiChat is attributed to two key factors. First, we develop a multi-agent framework to guide LLMs in generating the code for counterfactual constraints. The prompt is specifically tailored for why-not queries rather than general queries, guiding the LLM to generate a minimal amount of code under a narrowed task scope. Second, we incorporate various predefined functions to prevent LLMs from developing explanatory techniques from scratch for other queries. This approach is more robust than code generation, as demonstrated by the higher correctness rates in the first four queries in Table \ref{tab:comparison-gpt4o-mini}. More importantly, code generation is currently unable to develop more advanced explanatory techniques necessary for diagnosing queries and sensitivity queries. It is observed that code generation only produces heuristic techniques to tackle these queries, which results in suboptimal conclusions. Therefore, predefined functions are indispensable at the current stage. These factors contributing to the high accuracy will be discussed in more detail in the ablation studies (Section \ref{sec:ablation}).

\subsubsection{Failure Analysis} 
Despite achieving high accuracy, the LLM may occasionally fail to match a query to the appropriate function or generate code that deviates from the user's intended purpose. To better understand these model failures, we categorize errors into three types: \textit{syntax errors}, \textit{classification errors}, and \textit{logic errors}.

\textit{Syntax errors} include code execution failures and invalid function calls. These issues often arise from the ambiguity of natural language, which may mislead the LLM into generating function arguments or code that are not executable. Practitioners may describe the same model component using different terminology, while only one executable representation is valid in the model. This challenge is further amplified when components are indexed across multiple dimensions. Users may specify only a subset of dimensions, list them in an incorrect order, or omit them entirely. For example, \texttt{pc["max", :]} denotes the maximum production capacities for all facilities, yet practitioners might refer to it verbally as \textit{max output limits}, frequently omitting the qualifier \textit{for all facilities}. The LLMs might hallucinate by providing indices that do not exist.

\textit{Classification errors} occur when the explanation strategy selected by the LLM does not align with the human-annotated query type, such as misclassifying a \textit{why-not} query as a \textit{what-if} query.

\textit{Logic errors} arise in two scenarios. First, during code generation, the LLM may fail to represent a counterfactual scenario correctly using constraints. Second, the component names or indices generated by the LLM may exist in the model as valid function arguments, i.e., not a syntax error, but fail to accurately capture the user’s intent.

With this categorization in mind, the breakdown of errors using GPT-4.1 and o3 are shown in Table \ref{tab:error-types}. The proportions of syntax and classification errors in o3, 10. 0\% and 25. 0\%, respectively,  are notably lower compared to GPT-4.1, which exhibits 46.8\% syntax errors and 31.3\% classification errors. This indicates that o3 adheres more reliably to the syntax guidance and few-shot query demonstrations provided in the prompts. As a result, the proportion of logic errors becomes dominant in o3 (65.0\%) compared to GPT-4.1 (21.9\%). 



\begin{table}[ht]
\caption{Proportion Of Error Types}\label{tab:error-types}
\begin{tabular}{@{\extracolsep\fill}lcc}
\toprule
Error Type & GPT-4.1 (\%) & o3 (\%) \\
\midrule
Syntax Error         & $46.8$ & $10.0$ \\
Classification Error & $31.3$ & $25.0$ \\
Logic Error          & $21.9$ & $65.0$ \\
\bottomrule
\end{tabular}
\end{table}

\subsubsection{Ablation Studies}\label{sec:ablation}
To evaluate the impact of the proposed multi-agent framework, we perform ablation studies by removing the predefined functions, the syntax reminders, and the Illustrator, respectively. The results are shown in Table \ref{tab:unified_ablation} for the top two best performing models, GPT-4.1 and o3. 

When the predefined functions are removed in Table \ref{tab:unified_ablation}, OptiChat must rely entirely on code generation, i.e., the Programmer agent, to answer all queries. We observe a substantial drop in accuracy for both models, especially on the diagnosing and sensitivity queries. This indicates that the LLMs struggle to write code for retrieving the IIS or the dual variables of a constraint, tasks that demand deeper optimization expertise than the retrieval and what-if queries. Although both models suffer accuracy losses, o3 outperforms GPT-4.1, consistent with the claim that o3 is a superior reasoning model that excels at code generation. We also note a slight increase in response time without the predefined functions, reflecting the additional time the LLMs spend generating code versus leveraging existing tools.

\begin{table}[ht]
\caption{Ablation study results including experiments without predefined functions, syntax reminders, and illustrator.}\label{tab:unified_ablation}
\begin{tabular}{@{\extracolsep\fill}llcc|cc}
\toprule
 &  & \multicolumn{2}{c}{Accuracy (\%)} & \multicolumn{2}{c}{Time (min)} \\
\cmidrule{3-4} \cmidrule{5-6}
Query & Setting & GPT-4.1 & o3 & GPT-4.1 & o3 \\
\midrule
Diagnosing 
& Main & 89.7 & 89.7 & 0.1 & 0.6 \\
& w/o Predefined Functions & 0.0 & 33.3 & 0.2 & 1.5 \\
& w/o Syntax Reminders     & 41.0 & 87.2 & 0.1 & 0.5 \\
& w/o Illustrator           & 74.4 & 84.6 & 0.1 & 1.0 \\
\midrule
Retrieval  
& Main & 94.9 & 97.4 & 0.1 & 0.3 \\
& w/o Predefined Functions & 56.4 & 84.6 & 0.2 & 0.5 \\
& w/o Syntax Reminders     & 87.2 & 97.4 & 0.1 & 0.3 \\
& w/o Illustrator           & 74.4 & 87.2 & 0.1 & 0.7 \\
\midrule
Sensitivity 
& Main & 94.4 & 94.4 & 0.1 & 0.5 \\
& w/o Predefined Functions & 0.0 & 0.0 & 0.3 & 0.9 \\
& w/o Syntax Reminders     & 72.2 & 94.4 & 0.1 & 0.4 \\
& w/o Illustrator           & 83.3 & 83.3 & 0.1 & 0.8 \\
\midrule
What-if     
& Main & 84.6 & 87.2 & 0.1 & 0.4 \\
& w/o Predefined Functions & 82.1 & 92.3 & 0.2 & 0.9 \\
& w/o Syntax Reminders     & 66.7 & 84.6 & 0.1 & 0.3 \\
& w/o Illustrator           & 76.9 & 84.6 & 0.2 & 0.7 \\
\midrule
Why-not     
& Main & 62.2 & 78.4 & 0.3 & 0.9 \\
& w/o Illustrator           & 62.2 & 70.3 & 0.4 & 1.9 \\
\bottomrule
\end{tabular}
\end{table}

The syntax Reminder provides guidance for retrieving the correct arguments for predefined functions. When the Reminder is disabled, as shown in Table \ref{tab:unified_ablation}, we observe a slight accuracy drop in the o3 model and a significant drop in GPT-4.1. This further confirms the stronger reasoning capabilities of o3, while the weaker model tends to hallucinate without the syntax Reminder. The difference in response time is marginal, with or without the syntax Reminder.

The Illustrator agent extracts model components and their descriptions into a lookup table, providing essential context for answering various queries. This context is crucial for OptiChat’s performance. As shown in Table \ref{tab:unified_ablation}, both models experience a notable drop in accuracy without the Illustrator. Moreover, the response time increases significantly because the models must reprocess the original code to answer follow-up queries, rather than leveraging the pre-extracted context. 

\begin{figure}
\includegraphics[width=\linewidth]{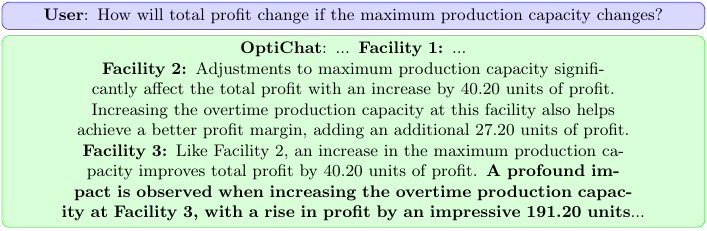}
\caption{A conversation with OptiChat. a, OptiChat helps practitioners identify sensitive parameters.}\label{fig:InsightSA}
\end{figure}

\begin{figure}
\includegraphics[width=\linewidth]{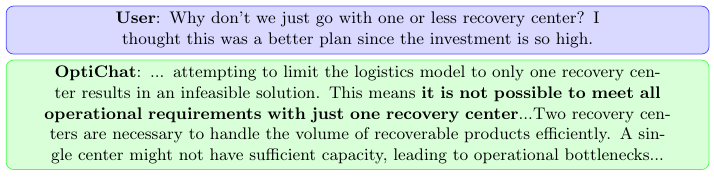}
\caption{A conversation with OptiChat. b, OptiChat helps practitioners embrace unanticipated decisions. }\label{fig:InsightCF}
\end{figure}

\subsubsection{Qualitative Assessment of Query Responses}
We showcase some insightful answers produced by OptiChat. The models used for the demonstration are built for supply chain management, which decides the optimal production, storage, and transportation of goods across production facilities, distribution centers, and markets. The maximum capacities of the normal and the overtime production often fluctuate due to factors such as raw material availability and labor engagement. Consider a case where a manufacturer is informed about an increase in raw materials and plans to expand maximum production capacities at certain facilities. As suggested in Figure \ref{fig:InsightSA}, the optimal profit is highly sensitive to the maximum overtime production capacity at the third facility, even though overtime production is commonly perceived as costly. Consequently, during the strategy development phase, it is crucial to prioritize the third facility over others and produce these additional items at the overtime stage. 

The second model takes into account recovery centers that manage re-manufacturing and recovery activities. This model suggests building two such recovery centers, despite the substantial investment cost typically associated with each. Suppose a situation where business managers express concerns regarding the financial burden and are inclined to construct fewer recovery centers. To reassure them, OptiChat justifies this decision by demonstrating that no feasible solution exists if forcing the number of recovery centers no greater than one in Figure \ref{fig:InsightCF}. Therefore, business managers can be convinced to deprecate outdated practices and appreciate the construction of recovery centers. 

\section{Conclusions and Future Work}\label{sec:conclusion}
In this paper, we propose OptiChat to address the common challenge that end users of optimization models are often not optimization experts. By leveraging the recent advancements in LLMs, OptiChat bypasses the need for inefficient back-and-forth coordination with optimization experts. Through straightforward natural language conversations with models, the practitioners can benefit from autonomous and instant responses. The integration of optimization tools with LLMs is demonstrated to be essential for providing a richer and more reliable context for analysis, rather than relying entirely on natural language processing to generate plausible explanations. 

Although OptiChat, at its current status, does not achieve perfect accuracy in executing tools or generating code, it is anticipated that this issue can be further mitigated by foreseeable advancements in LLM technology and targeted supervised fine-tuning specific to optimization tasks. One important step in this direction is the development of a larger and more diverse testing dataset to better evaluate and improve OptiChat's performance. Data augmentation techniques, such as prompting the LLM to rephrase the same questions, can be incorporated. Furthermore, users may exhibit varied preferences for answers and explanations, depending on their background and application context, especially in multi-turn conversations. To accommodate this, recent progress in reinforcement learning from human feedback (RLHF) offers a promising pathway. By training a reward model based on user preferences, we can better align OptiChat’s responses with the diverse expectations of its users. A more robust method can also be developed to evaluate users' satisfaction with OptiChat in multi-turn conversations. Last but not least, more explainable optimization techniques, such as inherently interpretable policies in the form of decision trees \citep{Goerigk2023AModels,Bertsimas2021TheOptimization}, can be added to the OptiChat. Thanks to OptiChat’s modular design, incorporating new explanation strategies or toolsets can be achieved with minimal overhead. We believe that OptiChat, along with the new dataset and the insights into query classification, could serve as a catalyst in this understudied area and draw more attention from the optimization and the machine learning community.   

\begin{appendices}
\section{Dataset Summary}\label{sec:datasetsumary}
Table~\ref{tab:model-summary} summarizes the collection of optimization models used in our study by type, size, and application domain. The dataset spans 24 models in total with medium sizes. For context, NLP4LP \citep{ahmaditeshnizi2024}, a dataset in the similar field, contains 355 instances in total (18 MILP instances), with parameter statistics of 3 / 8 / 18 (Min./Avg./Max.).

\begin{table}[ht]
\caption{Summary of Optimization Models by Type, Size, and Domain.} \label{tab:model-summary}
\begin{tabular}{@{\extracolsep\fill}lc}
\toprule
Category & Count \\
\midrule
\# Total Models & 24 \\
\# LP & 10 \\
\# MILP & 14 \\
\midrule
\# Parameters (Min./Avg./Max.) & 21 / 1,264 / 25,432 \\
\# Variables (Min./Avg./Max.)  & 14 / 1,002 / 14,880 \\
\# Constraints (Min./Avg./Max.) & 10 / 321 / 2,672 \\
\midrule
Manufacturing \& Scheduling & 8 \\
Supply Chain \& Logistics & 8 \\
Chemical \& Process Engineering & 4 \\
Power \& Energy & 3 \\
Other & 1 \\
\bottomrule
\end{tabular}
\end{table}

\section{Qualitative Evaluation of Model Descriptions}\label{sec:eval_md}
The quality of the model descriptions was assessed by expert participants. Each participant was provided with optimization models and asked to review the underlying code before comparing their own understanding with the descriptions generated by OptiChat. The evaluation considered several key aspects: (i) accuracy, captured by the presence of obvious errors or deviations from the actual meaning of the model; (ii) breadth, reflecting whether the description covered the important components of the model rather than focusing only on a narrow subset and (iii) clarity, indicating whether the description was approachable and would be easy for non-expert readers to understand. These criteria together served as the evaluation metric for determining the overall quality of the model descriptions.

\newpage
\section{Prompts}\label{sec:prompts}
\begin{lstlisting}[style=pythonprompt]
component_interpretation_prompt = """You are an operations research expert and your role is to use PLAIN ENGLISH to interpret an optimization model written in Pyomo. The Pyomo code is given below:
-----
{code}
-----
Here are the name of {component_type} that need to be described
-----
{component_names}
-----
Your task is carefully inspect the code and write a description for each of the components. 
Then, generate a json file accordingly with the following format (STICK TO THIS FORMAT!)
{model_interpretation_json}
- description should be either physical meanings, intended use, or any other relevant information about the component.
- Generate the complete json file and don't omit anything.
- Use 'name' and 'description' as the keys, and provide the name and description of the component as the values."""

model_illustration_prompt = """You are an operations research expert and your role is to introduce an optimization model to non-experts, based on an abstract representation of the model in json format.
The json representation is given below:
-----
{json_representation}
-----
- Start with a brief introduction of the model, what the problem is about, who is using the model, and what the model is trying to achieve.
- Explain what decisions (variables) are to be made
- Explain what data or information (parameters) is already known
- Explain what constraints are imposed on the decisions
- Explain what the objective is, what is being optimized
The explanation must be coherent and easy to understand for the users who are experts in the filed for which this model is built but not in optimization."""

iis_inference_prompt = """You are an operations research expert and your role is to infer why an optimization model is infeasible, based on an abstract representation of the infeasible model in json format.
Particularly, your team has identified the Irreducible Infeasible Subset (IIS) of the model, which is given below:
-----
{iis_info}
-----
To understand what the parameters and the constraints mean, the json representation is given below for your reference:
-----
{json_representation}
-----
- Introduce to the user what constraints are potentially causing the infeasibility, and what parameters are involved in these constraints.
- Explain the relationship between the constraints and the parameters, and infer why the constraints are conflicting with each other.
- Provide inference by analyzing their physical meanings, and AVOID using jargon and symbols as much as possible but the explanation style must be formal. 
- Recommend some parameters that you believe can be adjusted to make the model feasible.
- Parameters recommended for adjustment MUST be changeable physically in practice. For example, molecular weight of a molecule is not changeable in practice.
- Assess the practical implications of the recommendations. For example, increasing the number of workers implies hiring more workers, which incurs additional costs."""

coordinator_prompt = """You're a coordinator in a team of optimization experts. The goal of the team is to help non-experts analyze an optimization problem. Your task is to choose the next expert to work on the problem based on the current situation. 
Here's the list of agents in your team:
-----
{agents}
-----
Considering the conversation, generate a json file with the following format: 
{{ "agent_name": "Name of the agent you want to call next", "task": "The task you want the agent to carry out" }} 
to identify the next agent to work on the problem, and also the task it has to carry out. 
- Only generate the json file, and don't generate any other text.
- DO NOT change the keys of the json file, only change the values. Keys are "agent_name" and "task".
- if you think the problem is solved, generate the json file below:
{{ "agent_name": "Explainer", "task": "DONE" }}"""

explainer_prompt = """You're an optimization expert who helps your team answer user queries in MARKDOWN format.
- The users are not experts in optimization, but they are experts in the filed for which this model is built.
- Provide a detailed explanation only when you believe the users need more context about optimization to understand your explanation.
- Otherwise, the explanation must be succinct and concise, because users may be distracted by too much information.
- If Operators and Programmers in your team have provided technical feedback, then you need to summarize the feedback because the user cannot see them."""

syntax_reminder_prompt = """You're an operator working on a pyomo model.
Your task is to identify the following arguments: 
- the component names that the user is interested in,
- the most appropriate function that can answer the user's query, 
- the model that the user is querying.
then call the predefined syntax_guidance function to generate syntax guidance.
----- Instruction to select the most appropriate function -----
you MUST select a function from ```{function_names}```, DO NOT make up your own function.
1. feasibility_restoration:
Use when: The model is infeasible and you need to find out the minimal change to specific [component name] for restoring feasibility.
Example: "How much should we adjust the [component name] to make the model feasible"
Example: "I believe changing [component name] is practical, by how much do I need to change in order to make the model feasible"
[component name] category: parameters. If only constraint name is provided in the query, you need to infer the parameters involved in the constraint.

2. components_retrival:
Use when: You need to know the current values or expressions of [component name] within the model.
Example: "What are the values of the [component name]"
Example: "How many [component name] are currently available"
[component name] category: sets, parameters, variables, constraints, or objectives.

3. sensitivity_analysis:
Use when: The model is feasible and you want to understand the impact of changing [component name] on the optimal objective value, **without specifying the extent of changes**.
Example: "How will the optimal profit change with the change in the [component name]" (didn't specify how much the change is)
Example: "How stable is the objective value in response to variations in the [component name]" (didn't specify how much the change is)
Example: "Will the optimal value be greatly affected if we have more [component name]" (didn't specify how much the change is)
[component name] category: parameters.

4. evaluate_modification:
Use when: The model is feasible and you want to understand the impact of changing [component name] on the optimal objective value, **by specifying the extent of changes**.
Example: "How will the optimal profit change with **a 10% increase** in the [component name]" (specified the change is **a 10% increase**)
Example: "How stable is the objective value in response to the modification that [component name] is **decreased by 20 units**" (specified the change is **decreased by 20 units**)
Example: "Will the optimal value be greatly affected if we have **two more** [component name]" (specified the change is **two more**)
[component name] category: parameters or variables.

5. external_tools:
Use when: User doubts the model's optimal solution and provides a counterexample, and you want to add new constraints to implement the counterexample.
Example: "Why is it not recommended to have [component name] lower than 400 in the optimal solution"
Example: "Why isn't [component name] and [component name] both used in the optimal scenario"
[component name] category: parameters or variables.
    
----- Instruction to determine the correct component name -----
The [component name] MUST be in a symbolic form, instead of its description.
Use the following dictionary to find the correct [component name] based on its description:
{component_name_meaning_pairs}

----- Instruction to find the queried model -----
In the form of 'model_integer', e.g. 'model_1'"""

# all necessary information has been provided 
operator_prompt = """You're an optimization expert who helps your team to access and interact with optimization models by internal tools.
Your task is to invoke the most appropriate tool correctly based on the user's query and system reminders."""

code_reminder_prompt = """{source_code}\n# YOUR CODE GOES HERE\n"""

programmer_prompt = """You're an optimization expert who helps your team to write pyomo code to answer users questions, such as
- write code snippet to revise the model, only when the user doubts the model's optimal solution and provides a counterexample
- write code snippet to print out the information useful for answering the user's question
Output Format:
==========
```python
YOUR CODE SNIPPET
```
==========
Here are some example questions and their answer codes:
----- EXAMPLE 1 -----
Question: Why is it not recommended to use just one supplier for roastery 2?

Answer Code:
```python
# user is actually interested in the case that only one supplier can supply roastery 2 and does not believe the optimal solution.
model.force_one_supplier = ConstraintList()
model.force_one_supplier.add(sum(model.z[s,'roastery2'] for s in model.suppliers) <= 1)
for s in model.suppliers:
    model.force_one_supplier.add(model.x[s,'roastery2'] <= model.capacity_in_supplier[s] * model.z[s, 'roastery2'])
    from pyomo.environ import SolverFactory, TerminationCondition
    
# standard code to solve the model. Don't change this code if you need to solve a mode.
solver = SolverFactory('gurobi')  # only gurobi is available in env
solver.options['TimeLimit'] = 300  # 5min time limit
results = solver.solve(model, tee=False)  # tee must be False to suppress solver output, otherwise the output is overwhelming
print("Solver Status: ", results.solver.status)
print("Termination Condition: ", results.solver.termination_condition)
# always check the termination condition and optimal objective value first
if results.solver.termination_condition == TerminationCondition.optimal:
    from pyomo.environ import Objective
    from pyomo.environ import value
    for obj_name, obj in model.component_map(Objective).items():
        print('Optimal Objective Value: ', value(obj))
else:
    print("Model is infeasible or unbounded, no optimal objective value is available.")
    
# I print out the new optimal objective value so that you can tell the user how the objective value changes if only one supplier supplies roastery 2.
print('If forcing only one supplier to supply roastery 2, the optimal objective value will become: ', model.obj())
```

----- EXAMPLE 2 -----
Question: Why is it not recommended to have production cost larger than transportation cost in the optimal setting?

Answer Code:
```python
# user does not believe the optimal solution obtained when production cost smaller than transportation cost.
# so we force production cost to be less than transportation cost to see what will happen.
model.counter_example = ConstraintList()
model.counter_example.add(model.production <= model.transportation)
    
# standard code to solve the model. Don't change this code if you need to solve a mode.
solver = SolverFactory('gurobi')  # only gurobi is available in env
solver.options['TimeLimit'] = 300  # 5min time limit
results = solver.solve(model, tee=False)  # tee must be False to suppress solver output, otherwise the output is overwhelming
print("Solver Status: ", results.solver.status)
print("Termination Condition: ", results.solver.termination_condition)
# always check the termination condition and optimal objective value first
if results.solver.termination_condition == TerminationCondition.optimal:
    from pyomo.environ import Objective
    from pyomo.environ import value
    for obj_name, obj in model.component_map(Objective).items():
        print('Optimal Objective Value: ', value(obj))
else:
    print("Model is infeasible or unbounded, no optimal objective value is available.")
    
# I print out the new optimal objective value so that you can tell the user how the objective value changes.
print('If forcing production cost be smaller than transportation cost, the optimal objective value will become: ', model.obj())
```
- Code reminder has provided you with the source code of the pyomo model
- Your written code will be added to the lines with substring: "# YOUR CODE GOES HERE" So, you don't need to repeat the source code that has already been provided by Code reminder.
- The standard code for re-solving the model has been given in the examples, 
So, you MUST use the standard code to re-solve the model to avoid undesired execution errors and long execution result.
- Your written code should be accompanied by comments to explain the purpose of the code.
- Evaluator will execute the new code for you and read the execution result.
So, you MUST print out the model information that you believe is necessary for the user's question."""

evaluator_prompt = """You're an optimization expert who helps your team to review pyomo code,
based on the execution result of the code provided by the programmer.
Is the code bug-free and valid to answer the user's query?
Generate the following json file if you accept the code, and provide your own comment.{{ "decision": "accept", "comment": "your own comment" }}
Generate the following json file if you reject the code, and provide your own comment. {{ "decision": "reject", "comment": "your own comment" }}
- Only generate the json file, and don't generate any other text.
- Use 'decision' and 'comment' as the keys, 
- choose 'accept' or 'reject' for the decision, and provide your own comment. 
- Note that infeasibility caused by the new constraints may be acceptable. 
This is because programmers are trying to create a counterfactual example that the user is interested in, and this counterfactual example may be infeasible in nature."""

\end{lstlisting}
\newpage
\section{Workflow Example}\label{sec:workflow_example}
\begin{figure}[ht]
\centering
\includegraphics[width=0.8\linewidth]{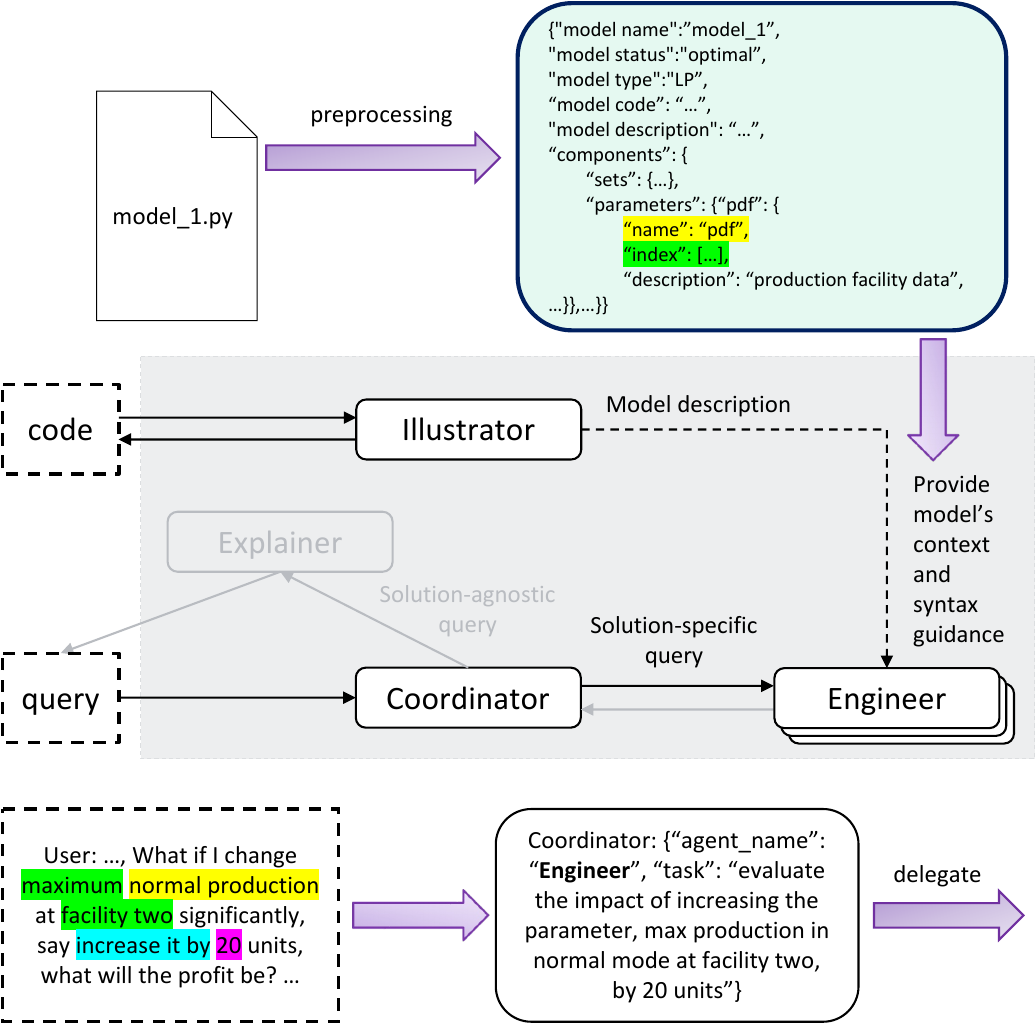}
\caption{An Illustrative example of the what-if query answered by the workflow (a) }\label{fig:workflow_example1}
\end{figure}

\begin{figure}[ht]
\centering
\includegraphics[width=0.8\linewidth]{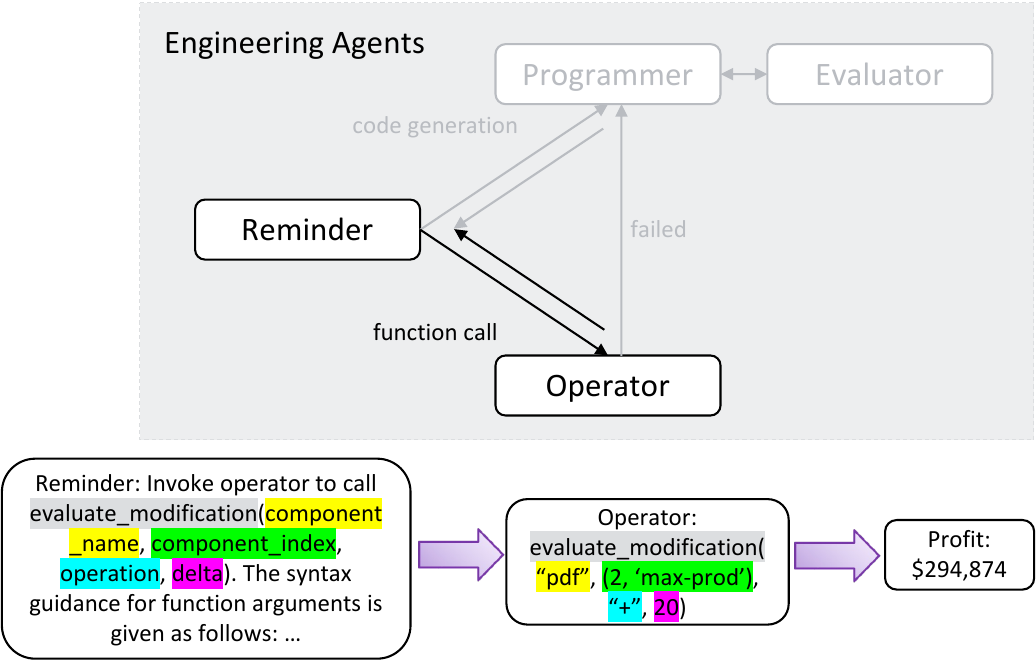}
\caption{An Illustrative example of the what-if query answered by the workflow (b) }\label{fig:workflow_example2}
\end{figure}

\begin{figure}[ht]
\centering
\includegraphics[width=0.8\linewidth]{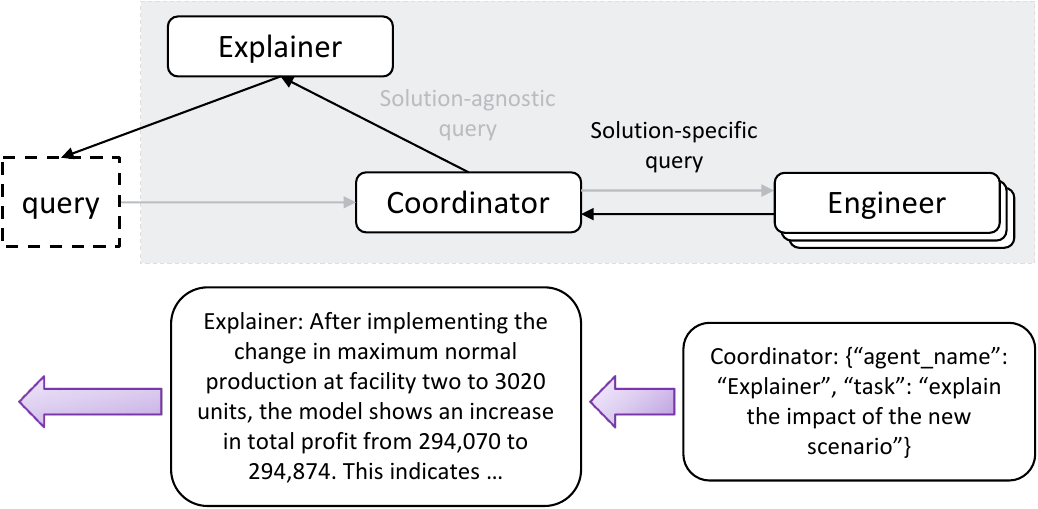}
\caption{An Illustrative example of the what-if query answered by the workflow (c)} \label{fig:workflow_example3}
\end{figure}

Figure~\ref{fig:workflow_example1}, Figure~\ref{fig:workflow_example2}, and Figure~\ref{fig:workflow_example3} show the workflow using an example query: ``what if I change maximum normal production at facility two significantly, say increase it by 20 units, what will the profit be?'' This query is first classified as solution-specific and delegated from the Coordinator to the Engineer. Among the Engineer's sub-agents, the reminder identifies that the appropriate explanation strategy for this what-if query is to evaluate the modification. The model information, previously preprocessed by the Illustrator, 
enables the reminder to correctly associate the term ``normal production'' with the parameter acronym ``pdf''. Based on this, the reminder invokes a predefined function to generate syntax guidance tailored to the ``evaluate modification'' function and the ``pdf'' component, capturing details such as the dimension and pattern of its indexes. This syntax guidance, along with the query, is then passed to the Operator, who formally invokes ``evaluate modification'' with the correct function arguments and computes the solution. Finally, the Explainer communicates the result to the user with context-aware natural language.

\newpage
\section{Example Conversation On User Interface}\label{sec:userconversation}
\begin{figure}[t]
\centering
\includegraphics[width=\linewidth]{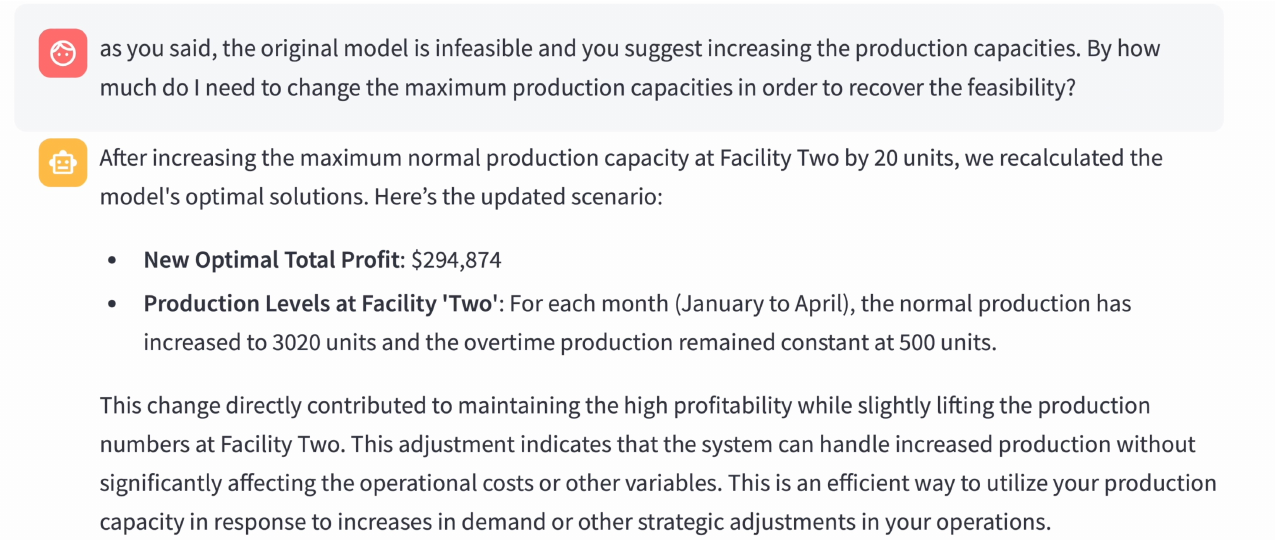}
\caption{A conversation example of the diagnosing query answered by OptiChat}\label{fig:diagnose_infeas}
\end{figure}

\begin{figure}[t]
\centering
\includegraphics[width=\linewidth]{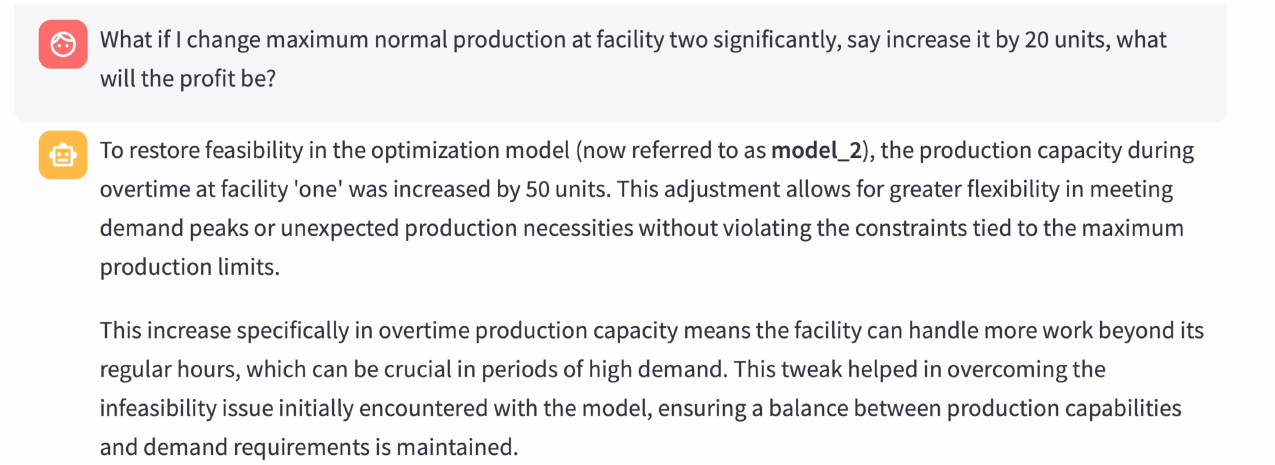}
\caption{A conversation example of the what-if query answered by OptiChat}\label{fig:eval_mod}
\end{figure}

Figure~\ref{fig:diagnose_infeas} and Figure~\ref{fig:eval_mod} present two representative conversations for the diagnosing and what-if queries.

\end{appendices}

\newpage
\bibliography{sn-bibliography}
\end{document}